\documentstyle[aps,prd,preprint,epsfig]{revtex}

\newcommand{\be}{\begin{equation}}
\newcommand{\ee}{\end{equation}}
\newcommand{\bb}{\begin{thebibliography}}
\newcommand{\eb}{\end{thebibliography}}
\newcommand{\bi}[1]{\bibitem{#1}}

\begin {document}

\title{Ergodic Properties of Classical SU(2) Lattice Gauge Theory}

\author{J. Bolte$^a$, B. M\"uller$^{b,c}$, and 
A. Sch\"afer$^b$ \\
$^a$ Abteilung Theoretische Physik, Universit\"at Ulm, D-89069 Ulm \\
$^b$ Institut f\"ur Theoretische Physik, Universit\"at Regensburg, 
D-93040 Regensburg \\
$^c$ Department of Physics, Duke University, Durham, NC 27708}

\date{\today}

\maketitle

\begin{abstract}
We investigate the relationship between the Lyapunov exponents of
periodic trajectories, the average and fluctuations of Lyapunov exponents
of ergodic trajectories, and the ergodic autocorrelation time for 
the two-dimensional hyperbola billiard.  We then study the fluctuation
properties of the ergodic Lyapunov spectrum of classical SU(2) gauge theory 
on a lattice. Our results are consistent with the notion that this system 
is globally hyperbolic. Among the many powerful theorems applicable to
such systems, we discuss one relating to the fluctuations in the entropy 
growth rate.
\end{abstract}


\section{Introduction}
Extensive experimental efforts are under way at Brookhaven National
Laboratory and CERN to produce and investigate the new deconfined, 
chirally symmetric high-temperature phase of QCD, usually called the 
quark-gluon-plasma (QGP). While the very high energy densities generated 
in high-energy nuclear collisions virtually guarantee that some new state 
of matter is reached, there are still important unresolved theoretical 
problems relating to the description of this state. One missing, critical 
ingredient is a non-perturbative approach to dynamical QCD processes far 
from thermodynamical equilibrium.

The study of non-equilibrium dynamics of relativistic quantum fields is
currently an active area of theoretical research \cite{NEQF}.
Approaches that go beyond perturbation theory include descriptions in
terms of probabilistic transport equations \cite{VBE}, and
deterministic or stochastic classical equations for the infrared
degrees of freedom of the quantum fields \cite{GM97,ASY99}. 
In the special, but important
case of non-Abelian gauge theories, the extreme infrared limit has
been long known to correspond to a dynamical system exhibiting classical
as well as quantum chaos \cite{Mat,Sav}.  Several years ago, this result
was extended to spatially varying, lattice regulated Yang-Mills fields
by numerical calculation of the maximal Lyapunov exponents and the 
complete ergodic Lyapunov spectrum of classical SU(2) gauge theory
\cite{ba,gong1,gong2}. The most intriguing results with implications for 
relativistic heavy ion physics are:
\begin{enumerate} 
\item The ergodic Lyapunov spectrum looks exactly as expected for 
a globally hyperbolic system.
\item The largest Lyapunov exponent appears to be related to the plasmon
damping rate as predicted by high temperature perturbation theory
\cite{Biro}.
\item The magnitude of the maximal Lyapunov exponent for SU(3) indicates 
a rapid thermalization of gluons in heavy-ion collisions.
\end{enumerate}
These results suggest the extension of this approach to a systematic 
semi-classical description of the dynamics of Yang-Mills field theories. 
The success of such an approach will ultimately depend on one's ability
to find practical methods for the application of periodic orbit theory 
to systems with many degrees of freedom. We discuss a very first 
step in this direction.

We present results of an investigation of the relation between the 
Lyapunov exponents of periodic and ergodic orbits. Periodic orbit theory, 
in the framework of the thermodynamic formalism, makes detailed predictions 
for the statistical properties of Lyapunov exponents of generic orbits 
in Anosov systems, but few studies of these relations appear to have 
been made for specific chaotic, non-linear dynamical systems. This 
motivated our numerical study of the relation between the Lyapunov 
exponents of periodic orbits and generic trajectories in a system for 
which the Lyapunov exponents of periodic orbits (henceforth simply 
called ``periodic Lyapunov exponents'') are known for all orbits below 
a certain period: the two-dimensional hyperbola billiard \cite{jb}. 
For this system, therefore, powerful mean value theorems can be invoked 
to predict analytical relations which can be checked numerically. 
Below, we present our numerical results confirming the general connection 
between the Lyapunov exponents for ergodic and periodic orbits as well as 
for their fluctuations. 

We then dicuss the corresponding properties of the Lyapunov exponents of 
ergodic trajectories for classical SU(2) Yang-Mills theory on a 
three-dimensional lattice. The observed similarities suggest that this 
system is also globally hyperbolic and could, in principle, be treated
within the framework of periodic orbit theory. Our conjecture yields
a prediction for the fluctuation properties of the ergodic Lyapunov 
exponents which is verified numerically. On the basis of the relation
between these fluctuations and the fluctuations of the entropy growth rate
we obtain a prediction of the magnitude of entropy fluctuations as a 
function of space-time volume. We find that for the conditions occurring 
in high energy nuclear collisions these fluctuations are expected to be 
very small, in agreement with observations.

We emphasize that it is presently impossible to predict how 
far this approach will carry toward a description of the dynamics of
non-equilibrium processes in QCD. Classical Yang-Mills equations can only 
be used to estimate a very limited number of dynamical parameters of the
QGP, namely those which have a well-defined classical limit, such as the
logarithmic entropy growth rate, $d(\ln S)/ dt$, but not quantities such
as the energy or entropy density. Our analysis is only relevant to the 
fluctuation properties of such, essentially ``classical'' quantities. 
However, we hasten to point out that, independently of the specific 
application considered here, an improved understanding of the connection 
between quantum field theory and periodic orbit theory is of fundamental 
theoretical relevance for non-linear dynamics in general. To our knowledge
for the first time, we propose a general relationship between the mean 
periodic Lyapunov exponents of a dynamical system, its mean ergodic Lyapunov 
exponents, and the ergodic autocorrelation time. This general relationship
makes it possible to extract important new information for any
higher-dimensional system for which the explicit construction of the
periodic orbits is practically not feasible.

The basic assumption underlying periodic orbit theory is that the periodic 
orbits sample the phase space of a non-linear dynamical system in such a 
manner that its averaged properties can be systematically reconstructed
from the properties of the periodic orbits. For each such orbit there
is a spectrum of characteristic Lyapunov exponents that describe how fast 
the separation between neighboring orbits increases with time. While periodic 
orbit theory is an extremely powerful tool, its range
of applicability is strongly limited by the difficulties encountered 
in determining the complete set of periodic orbits.
For any field theory with its potentially infinitely many degrees of freedom, 
the task of numerically constructing the periodic orbits looks hopeless. 
It is, however, relatively easy to obtain ergodic Lyapunov exponents by
numerical integration of the equations of motion \cite{ba}. Since it seems 
plausible that every ergodic trajectory eventually comes close to any periodic 
orbit, any infinite ergodic orbit should sample all periodic orbits. Thus it
appears as a natural conjecture that the average properties of ergodic
Lyapunov exponents and the average properties of periodic Lyapunov exponents 
should be related. It is this relationship that we want to discuss in the
following.

\section{General Relations}

Before we investigate and confirm the relationship between ergodic and 
periodic orbits for a simple but non-trivial system for which
all periodic Lyapunov exponents (up to a certain period) are known, 
namely, the two-dimensional hyperbola billiard \cite{jb}, we review
some general relations between Lyapunov exponents of periodic and
generic trajectories.  In the next section, we will compare these 
analytic predictions for the properties of the ergodic Lyapunov exponents 
$\lambda_{\rm r}$ with those obtained by numerical integration of a randomly 
chosen ergodic trajectory $\vec x(t) = \vec x_0(t)+\delta\vec x(t)$:
\be
\lambda_{\rm r} =
\lim_{\delta \vec x(0)\rightarrow 0}\lim_{t\rightarrow \infty} {1\over t} 
\ln {\vert\delta \vec x(t)\vert\over \vert\delta \vec x(0)\vert}\ ,
\label{eq1}
\ee
where the index $r$ indicates the random starting point.
(We remind the reader that for a fully ergodic system this yields the
maximal ergodic Lyapunov exponent, which for $d=2$ degrees of freedom is
the unique positive exponent.)

In a Hamiltonian hyperbolic dynamical system with $d$ degrees of freedom
ergodicity implies that the sum of its $d-1$ positive ergodic Lyapunov 
exponents can also be obtained as the ergodic mean of the local expansion rate,
\be
\lim_{t\rightarrow \infty} h_{\rm r}(t) \equiv
\lim_{t\rightarrow \infty} {1\over t} \int_0^t \chi(\vec x(t'))\ dt'
=\sum_{j=1}^{d-1}\lambda_{r,j} = h_{\rm KS}\ .
\label{eq2}
\ee
Here $h_{\rm KS}$ denotes the Kolomogorov-Sinai entropy and
\be
\chi(\vec x(t)) = {d\over dt}\ln\det\left( 
{\partial\vec x(t) \over \partial\vec x(0)} \right)_{\rm expanding}
\label{eq3}
\ee
is the local rate of expansion along the trajectory $\vec x(t)$.
Due to the equidistribution of periodic orbits in phase space it is 
possible to evaluate the ergodic mean in (\ref{eq2}) by weighted sums 
over periodic orbits. In fact, for hyperbolic systems the thermodynamic 
formalism allows to express certain invariant measures on phase space
in terms of averages over periodic orbits, see, e.g., \cite{ParPol,ga}. 
One is in particular able to obtain a relation that establishes a direct 
connection between the positive ergodic Lyapunov exponents $\lambda_{r,j}$ 
and those of periodic orbits. Labelling periodic orbits by $\nu$, and 
denoting their periods and positive Lyapunov exponents by $T_{\nu}$ and 
$\lambda_{\nu,j}$, respectively, this relation reads
\be
\sum_{j=1}^{d-1}\lambda_{r,j} = \lim_{t\rightarrow\infty}
\frac{\sum_{t\leq T_\nu\leq t+\varepsilon} \left( \sum_{j=1}^{d-1}
\lambda_{\nu,j}\right) \exp\left( -\sum_{j=1}^{d-1}\lambda_{\nu,j}
T_\nu\right)} {\sum_{t\leq T_\nu\leq t+\varepsilon}
\exp\left( -\sum_{j=1}^{d-1}\lambda_{\nu,j}T_\nu\right)} \ ,
\label{eq4}
\ee
where $\varepsilon>0$ is arbitrary. Within the thermodynamic formalism the 
topological pressure $P(\beta)$ was introduced as a useful tool to analyze
invariant measures on phase space in terms of periodic orbits as, e.g., in 
(\ref{eq4}). This function can be expressed as 
\be
P(\beta) = \lim_{t\rightarrow\infty} {1\over t} \ln 
\sum_{t\leq T_\nu\leq t+\varepsilon} \exp\left( -\beta \sum_{j=1}^{d-1}
\lambda_{\nu,j}T_\nu\right)\ ,
\label{eq5}
\ee
and it is not difficult to derive from (\ref{eq5}) that  
$P(\beta)$ is monotonically decreasing and convex. The exponential 
proliferation of the number of periodic orbits immediately implies that 
$P(0)=h_{top}$ (topological entropy). Moreover, the arithmetic average of 
the sum of the positive periodic Lyapunov exponents is given by 
$\bar\lambda=-P'(0)$. The relation (\ref{eq4}) then follows from (\ref{eq2}) 
and from the non-trivial identity $-P'(1)=h_{\rm KS}$. One also concludes that 
the three quantities measuring a mean separation of neighboring trajectories 
are ordered in the following way: $\bar\lambda \geq h_{top} \geq h_{\rm KS}$. 
For further information see, e.g., \cite{ga}.

Our next goal is to investigate the fluctuations of the local rate of
expansion (\ref{eq3}), when integrated up to a sampling time $t_{\rm s}$, 
about its ergodic mean (\ref{eq2}). We recall that this quantity was
denoted as $h_{\rm r}(t_{\rm s})$ in (\ref{eq2}).
For (uniformly) hyperbolic dynamical systems one expects that observables 
sampled along ergodic trajectories up to time $t_{\rm s}$ show Gaussian 
fluctuations about their ergodic mean. Indeed, in many cases a central limit 
theorem holds true that also predicts the widths of these Gaussian to scale as
$t_{\rm s}^{-1/2}$ for large sampling times $t_{\rm s}$. More precisely, 
Waddington \cite{wa} has shown that for Anosov systems (i.e., fully 
hyperbolic systems on compact phase spaces) the difference
\be
\sqrt{t_{\rm s}}\left[ h_{\rm r}(t_{\rm s}) - h_{\rm KS} \right]
\label{eq7}
\ee
shows Gaussian fluctuations with variance $P''(1)$ in the limit 
$t_{\rm s}\rightarrow\infty$. This means that
\be
\Delta h_{\rm r}(t_{\rm s})\sim \sqrt{P''(1)/t_{\rm s}}\ ,\quad 
t_{\rm s}\rightarrow\infty\ .
\label{eq8}
\ee
According to (\ref{eq5}) the quantity $P''(1)$ can be 
expressed in terms of periodic orbit sums as
\be
P''(1) =  \lim_{t\to\infty} t 
\left[ \frac{\sum_{\nu}\left( \sum_{j}\lambda_{\nu,j}\right)^2
\exp\left( -\sum_{j}\lambda_{\nu,j}T_\nu\right)}
{\sum_{\nu}\exp\left( -\sum_{j}\lambda_{\nu,j}T_\nu\right)} -
\left( \frac{\sum_{\nu}\left(\sum_{j}\lambda_{\nu,j}\right)
\exp\left( -\sum_{j}\lambda_{\nu,j}T_\nu\right)}
{\sum_{\nu}\exp\left( -\sum_{j}\lambda_{\nu,j}T_\nu\right)}
\right)^2  \right] \ .
\label{eq9}
\ee
On the other hand, the variance of the distribution of the periodic Lyapunov 
exponents is related to $P''(0)$, since
\be
P''(0) =  \lim_{t\to\infty} t 
\left[ \frac{\sum_{t\leq T_\nu\leq t +\varepsilon}
\left( \sum_{j=1}^{d-1}\lambda_{\nu,j}\right)^2}
{\sum_{t\leq T_\nu\leq t +\varepsilon}1} -
\left( \frac{\sum_{t\leq T_\nu\leq t +\varepsilon}\left(
\sum_{j=1}^{d-1}\lambda_{\nu,j}\right)}
{\sum_{t \leq T_\nu\leq t +\varepsilon}1}\right)^2  \right] \ .
\label{eq10}
\ee

For the hyperbola billiard this variance was calculated numerically by Sieber 
\cite{jb}, who found Gaussian distributions of the positive Lyapunov 
exponents of periodic orbits with $N$ bounces off the boundary. For large $N$
the widths of these Gaussian scale like
\be
\tilde\sigma_N\sim {0.199 \over \sqrt{N}}\ .
\label{eq11}
\ee
Taking into account that the mean length of periodic orbits with $N$ bounces
scales as $\bar t_N\sim 2.027 N$ \cite{jb}, this yields a prediction for the
width of the distribution of periodic Lyapunov exponents expressed as a 
function of $t$ that scales as  
\be
\Delta\lambda_{\nu}(t)\sim {0.283 \over \sqrt{t}}
\label{eq12}
\ee
in the limit of long periodic orbits. One hence concludes that 
$P''(0)=0.08$.

The variance of the fluctuations (\ref{eq7}) can also be related to the 
autocorrelation function
\be
a (\tau) = \langle \chi(\vec x(\tau))\,\chi(\vec x(0)) \rangle - (h_{\rm KS})^2
\label{eq13}
\ee
of the local ergodic Lyapunov exponents, where $\langle\dots\rangle$ denotes 
a phase space average. In order to derive this connection one averages the 
square of (\ref{eq7}) over phase space, which then leads to
\be
t\,(\Delta h_{\rm r}(t))^2 = {1 \over t}\int_{-t}^{+t}(t-|\tau|)\,a (\tau)
\ d\tau\ .
\label{eq14}
\ee
A connection with the topological pressure can be established because
(\ref{eq7}) and (\ref{eq8}) imply that the autocorrelation function 
(\ref{eq13}) vanishes faster than $1/\tau$ as $\tau\to\infty$. One can 
therefore perform the limit $t\to\infty$ on both sides of (\ref{eq14}), 
yielding
\be
P''(1) = \lim_{t\rightarrow\infty} t\,(\Delta h_{\rm r}(t))^2 = 
\int_{-\infty}^{+\infty}a (\tau)\ d\tau\ .
\label{eq15}
\ee

Finally we want to discuss the probability for deviations of the sum of the 
positive ergodic Lyapunov exponents, sampled over time $t$, from its ergodic 
mean $h_{\rm KS}$. To this end let $p_t (h)$ denote the probability density 
for $h_{\rm r}(t)$ to have a value $h$. Waddington has shown \cite{wa} 
that for Anosov systems which are such that $P''(\beta)\neq 0$ for all 
$\beta$, this probability density has the form
\be
p_t(h) = f(h)\,\sqrt{t}\,\exp(-g(h)t)\ ,
\label{eq17}
\ee
where $f(h)$ is a complicated, though uniquely fixed function. Moreover, 
\be
g(h)= \inf_{\beta} \{ h\beta + P(\beta+1) \}
\label{eq18}
\ee
is a strictly convex, non-negative function with a unique minimum at the 
ergodic mean $h_{min}=h_{\rm KS}$, where $g(h_{\rm KS})=0$. This means 
that for large $t$ the probability of large deviations of $h_{\rm r}(t)$ 
from the ergodic mean is exponentially small.

\section{The Two-Dimensional Hyperbola Billiard}

We test the above statements in the two-dimensional hyperbola billiard,
for which all periodic orbits and their Lyapunov exponents are known
up to certain orbit period \cite{jb}. In order to be able to compare
our numerical results with the analytical predictions, which are based
on periodic orbits in a restricted length range, we have limited the 
motion into the arms of the hyperbola billiard, using the cut-off
$\vert x\vert, \vert y\vert \leq x_{\rm lim} = 10/\sqrt{2}$ and
reflecting the motion horizontally or vertically at the boundary. 
We have not studied the dependence of our results on $x_{\rm lim}$ in 
any systematic fashion, but a cursory exploration did not reveal a
significant dependence.

Our numerical result for the KS-entropy was obtained as 
$h_{\rm KS} \equiv \lambda_{\rm r} = 0.575$ by exploiting the relation
(\ref{eq1}) for the positive ergodic Lyapunov exponent. In \cite{jb} the 
arithmetic average of the periodic Lyapunov exponents and the topological 
entropy have been determined numerically as $\bar\lambda = 0.703$ and 
$h_{\rm top} = 0.5925$, respectively, so that the ordering 
$\bar\lambda \geq h_{\rm top} \geq h_{\rm KS}$ is respected. This provides 
a non-trivial test since the general theoretical statement has only been
proven for uniformly hyperbolic systems (Anosov systems) on compact phase 
spaces. The hyperbola billiard is only non-uniformly hyperbolic and, 
moreover, without the imposed cut-off its phase space fails to be compact. 

\begin{figure}[htb]
\centerline{\mbox{\epsfig{file=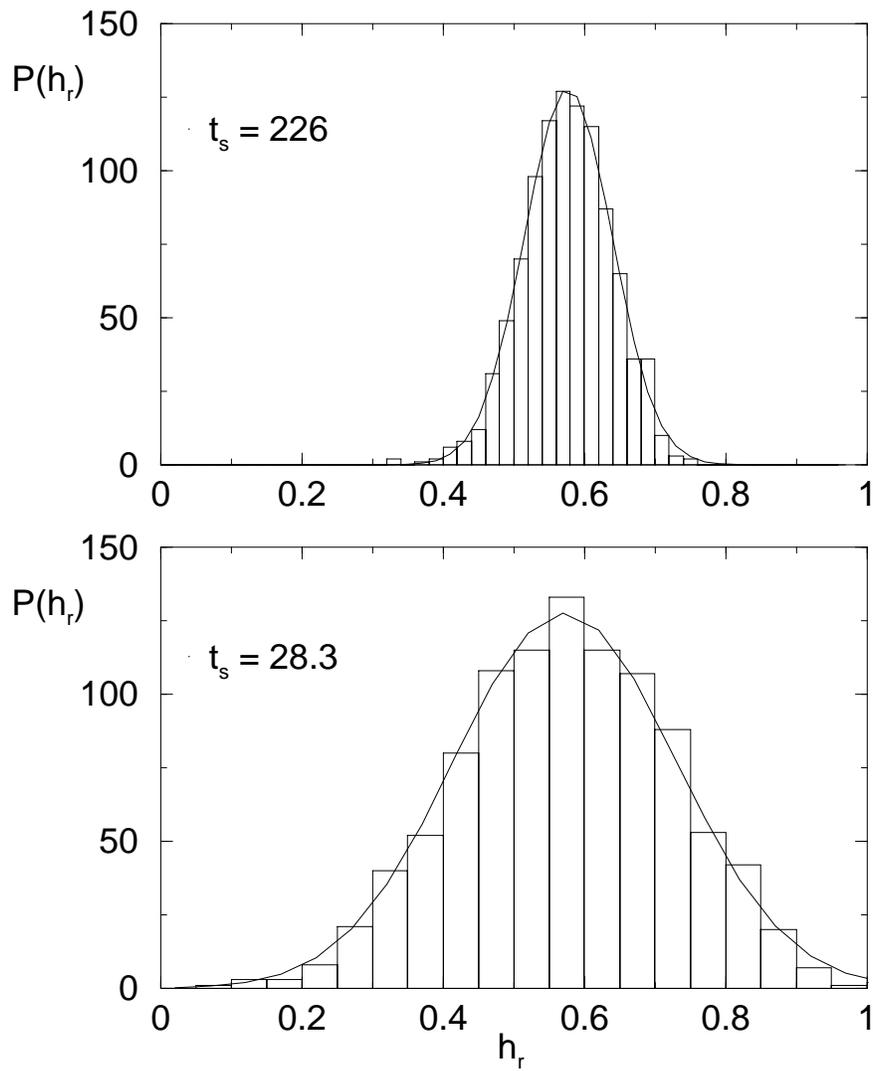,width=0.7\linewidth}}}
\bigskip
\caption{Distribution the calculated local Lyapunov exponents for ergodic 
trajectories of two different length $t_{\rm s}$ in the hyperbola billiard.}
\label{fig1}
\end{figure}

\begin{figure}[tbh]
\centerline{\mbox{\epsfig{file=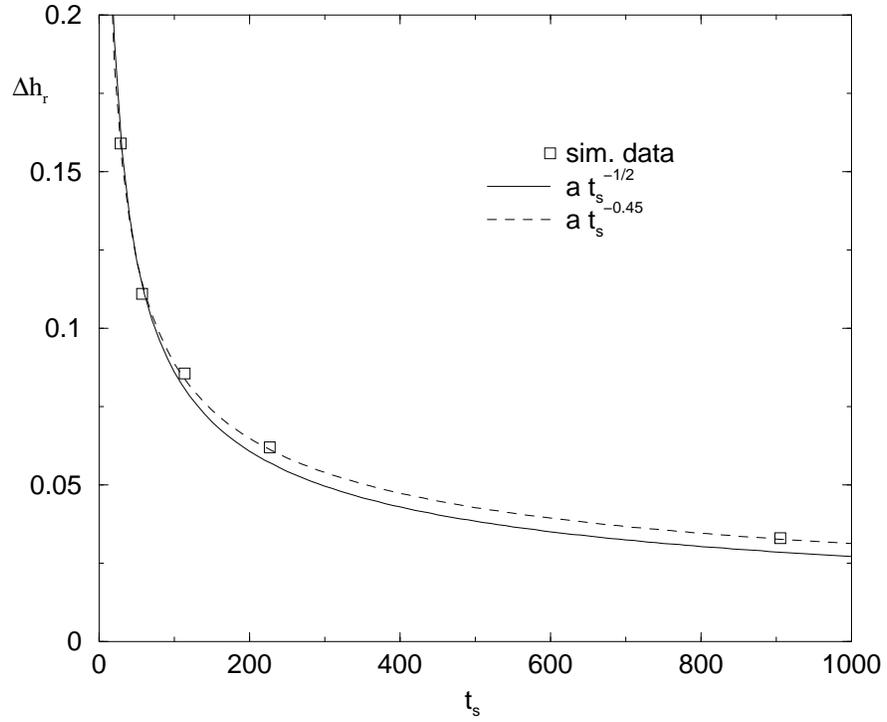,width=0.7\linewidth}}}
\bigskip
\caption{The widths of the Gaussians illustrated in Fig.~\protect\ref{fig1}
as a function of $t_{\rm s}$ together with a fits of the form 
$at_{\rm s}^{-1/2}$ (solid line) and $at_{\rm s}^{-b}$ with $b=0.45$ 
(dashed line).}
\label{fig2}
\end{figure}

For the (cut-off) hyperbola billiard we found that the distributions of 
the ergodic Lyapunov exponents (\ref{eq1}) that we determined numerically 
up to sampling times $t_{\rm s}$ are very well described by Gaussians, 
see Fig.~\ref{fig1}, if the sampling time is not too small 
($t_{\rm s} \gg 1$). For small sampling times, most of the phase space
divergence occurs during intervals $t_{\rm s}$ when the trajectory
reflects off the hyperbolic boundary, making the distribution of 
$h_{\rm r}(t_{\rm s}) \equiv \lambda_{\rm r}(t_{\rm s})$ strongly
non-Gaussian in the limit $t_{\rm s} \to 0$.  We made power-law fits 
of the form $a t_{\rm s}^{-b}$ to the dependence  of the widths of 
these Gaussians on $t_{\rm s}$. This gave the result (see Fig.~\ref{fig2}):
\be
\Delta h_{\rm r}(t_{\rm s})\approx 0.86\, t_{\rm s}^{-1/2}\ .
\label{eq6}
\ee

\begin{figure}[htb]
\centerline{\mbox{\epsfig{file=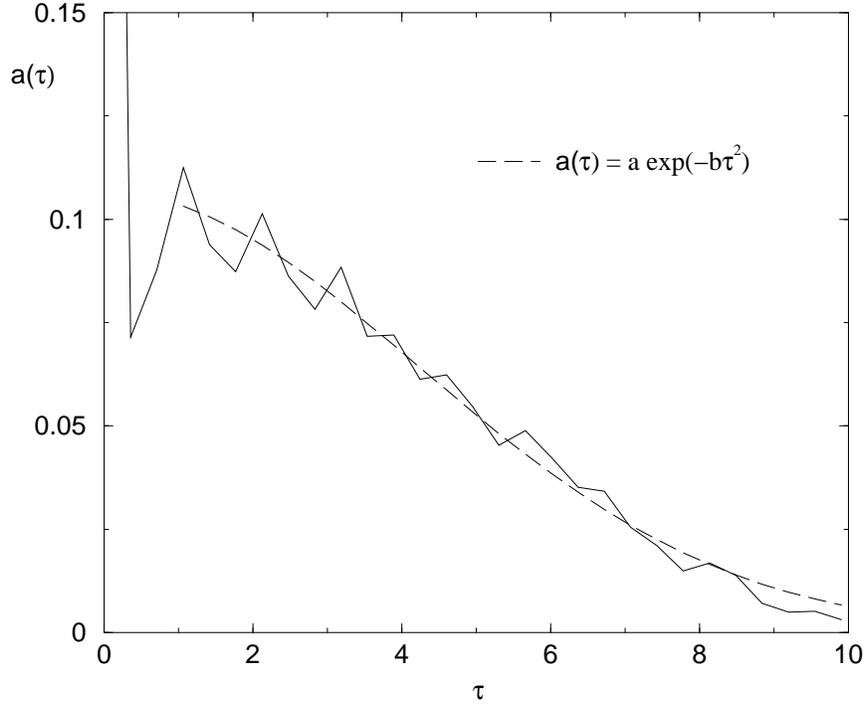,width=0.7\linewidth}}}
\bigskip
\caption{Temporal autocorrelation for the local Lyapunov exponents 
determined along an ergodic trajectory in the two-dimensional hyperbola 
billiard. The dashed line is a fit of the form $a\exp(-\tau^2/t_c^2)$
yielding $t_c\approx 6$.}
\label{fig3}
\end{figure}

We also determined the correlation function $a(\tau)$ for the hyperbola 
billiard by sampling $h_{\rm r}(t_{\rm s})$ for in small intervals
$t_{\rm s} = 1/(2\sqrt{2})$. The result is shown in Fig.~\ref{fig3}. 
Clearly, $a(\tau)$ falls off rapidly with a time constant of about
$t_c = 6$. Therefore, we can test the relation (\ref{eq14}) by
integrating the right-hand side numerically. For $t = 28.3$, 
corresponding to the lower plot in Fig.~\ref{fig1}, we obtain in this
way the prediction $\Delta h_{\rm r} = 0.197$ with an estimated numerical 
uncertainty of about 25\%. The value obtained from the Gaussian fit to 
the histogram in Fig.~\ref{fig1} is $\Delta h_{\rm r} = 0.159$. 
The quality of this agreement must be judged with the fact in mind
that the correlation function $a(\tau)$, as well as the distribution
$\chi(\vec x(t))$ are highly singular for the hyperbola billiard
in the limit $\tau\to 0$.

\section{The SU(2) gauge theory on a lattice}

Let us now turn to a comparison with results obtained for ergodic 
orbits in the classical SU(2) Yang-Mills theory regularized on a 
lattice. In \cite{gong2} the complete (positive) Lyapunov spectra were 
obtained for lattice volumes $L^3$ with $L=1,2,3$. We have extended 
these calculations to the lattices of size $L=4,6$. All our calculations
were performed for an average energy per plaquette $E_{\rm p}\approx 1.8$.
For sufficiently long trajectories and fixed energy per lattice site the 
Lyapunov spectrum has a unique shape, independent of the lattize size,
as shown in Fig.\ref{fig4}. Indeed, for a completely hyperbolic system, 
physical intuition requires that the Kolmogorov-Sinai entropy $-P'(1)$ 
is an extensive quantity. For this to be true, the sum over all positive 
Lyapunov exponents must scale like the lattice volume $L^3$ and the shape of 
the distribution of Lyapunov exponents must be independent of $L$. 
Figure \ref{fig4} confirms this expectation.

\begin{figure}[htb]
\centerline{\mbox{\epsfig{file=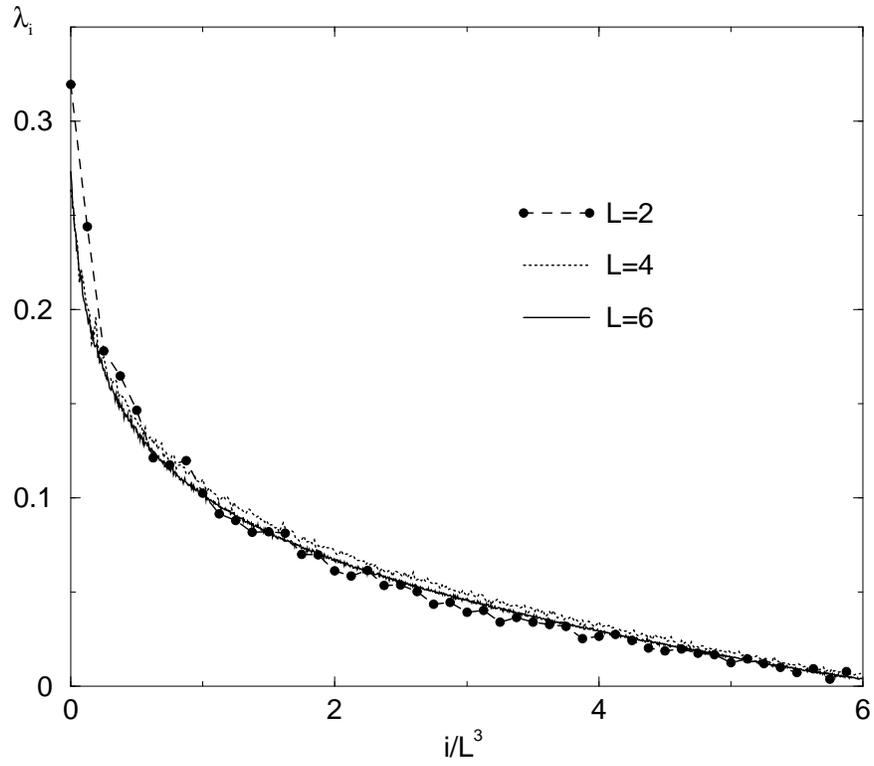,width=0.7\linewidth}}}
\bigskip
\caption{Distribution of numerically obtained ergodic Lyapunov
exponents for a classical SU(2) gauge theory on lattices of size
$L=2,4,6$. The index $i$ numbers the Lyapunov exponents and the
abscissa is scaled with $L^3$.}
\label{fig7}
\end{figure}

\begin{figure}[htb]
\centerline{\mbox{\epsfig{file=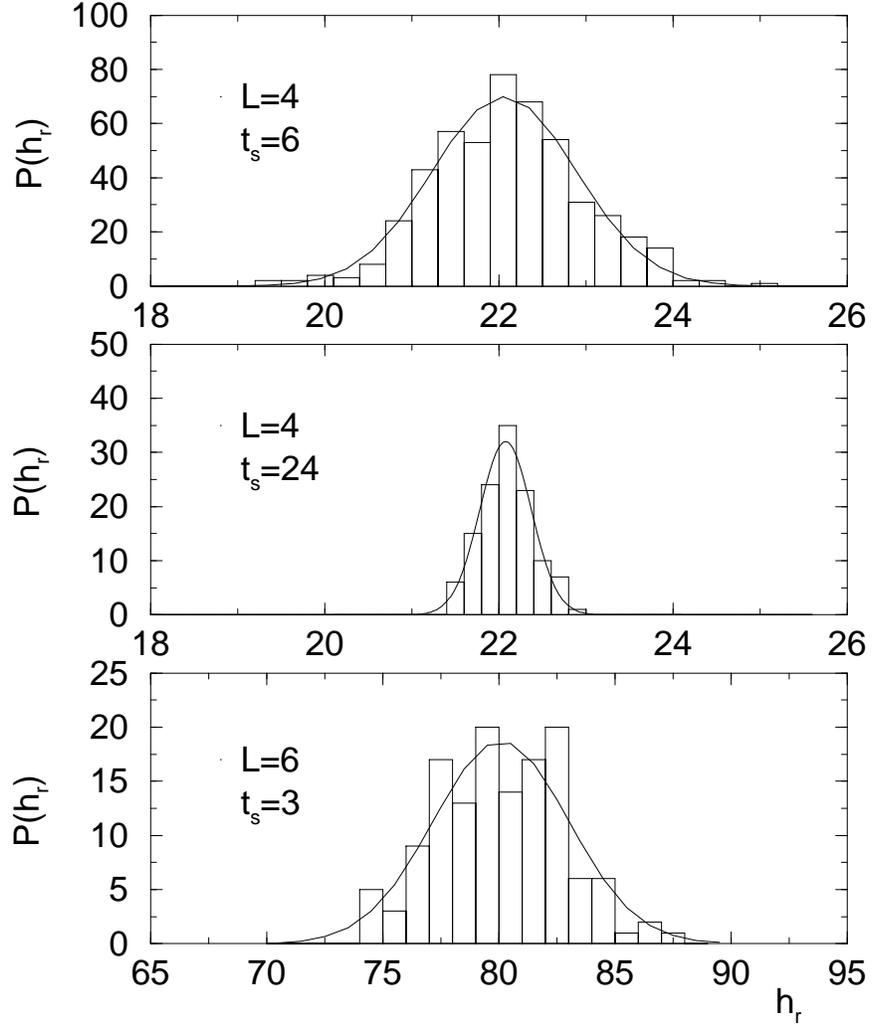,width=0.7\linewidth}}}
\bigskip
\caption{Distributions of the sum of the positive ergodic Lyapunov
exponents for different values of the trajectory length $t_{\rm s}$ for classical 
lattice SU(2) gauge theory on lattices of size $L=4,6$.}
\label{fig4}
\end{figure}

\begin{figure}[htb]
\centerline{\mbox{\epsfig{file=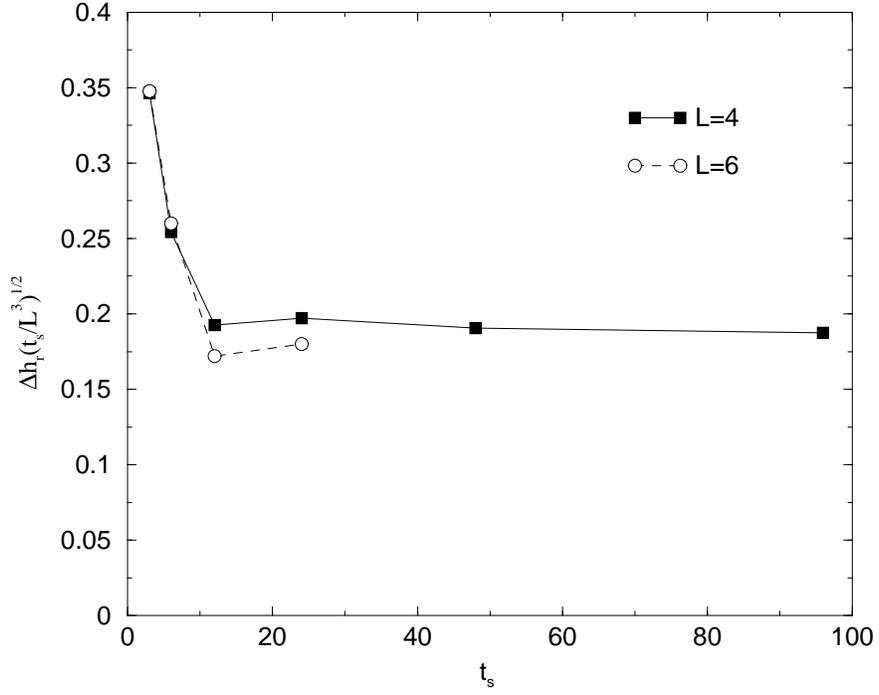,width=0.7\linewidth}}}
\bigskip
\caption{The widths of the Gaussians illustrated in Fig.~\protect\ref{fig4}
and scaled with $(t_{\rm s}/L^3)^{1/2}$, as a function of $t_{\rm s}$.}
\label{fig5}
\end{figure}

\begin{figure}[htb]
\centerline{\mbox{\epsfig{file=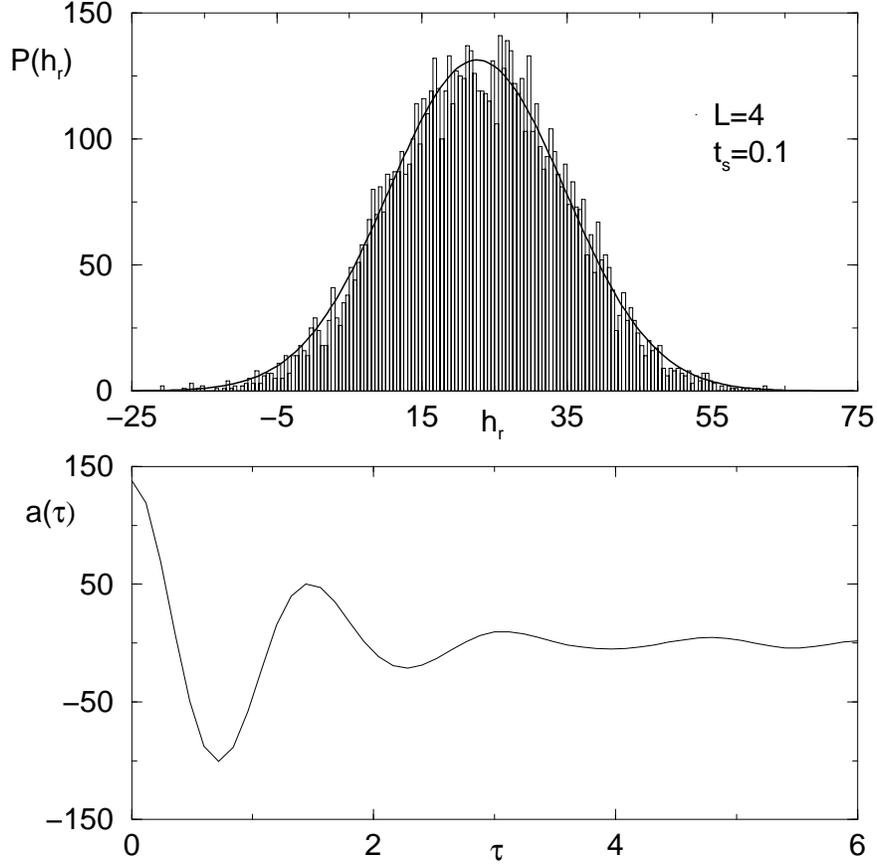,width=0.7\linewidth}}}
\bigskip
\caption{Top: The distribution of the sum of local expansion rates $h_{\rm r}$ 
for L=4 and a short sampling time $t_{\rm s}=0.1$, together with a 
Gaussian fit. Because the globally expanding phase space volume may
be locally contracting, the distribution has a small tail extending
to negative values of $h_{\rm r}$. This tail disappears (becomes
exponentially small) for large values of $t_{\rm s}$.
Bottom: The autocorrelation function $a(\tau)$ for this distribution.}
\label{fig6}
\end{figure}

In Fig.~\ref{fig5} we show distributions of the 
sum over positive Lyapunov exponents as a function of the 
length of the sampled ergodic trajectories (obtained as function of
the sampling time $t_{\rm s}$ on a single, very long trajectory). Obviously, 
the distributions are nicely fitted by Gaussians whose widths decrease 
like $1/\sqrt{t_{\rm s}}$ (see Fig.~\ref{fig6}). This behavior is identical 
to that of the two-dimensional hyperbolic system studied before 
(cf.~Fig.~\ref{fig1}). We also determined again the autocorrelation 
function $a(\tau)$ defined in (\ref{eq13}) by sampling the distribution
$p_t(h)$ with small time steps (see top part of Fig.~\ref{fig7}).  
For the $L=4$ lattice the result is shown in the lower part of 
Fig.~\ref{fig7}.  This allows us to test the relation (\ref{eq14}) 
connecting the $a(\tau)$ with the variance of the ergodic Lyapunov 
exponents.  Using (\ref{eq14}) we obtain the value $\Delta h_{\rm r}=0.88$ 
for $t_{\rm s}=6$, whereas the Gaussian fit to the sampled distribution 
shown in the top part of Fig.~\ref{fig5} is $\Delta h_{\rm r}=0.83$.  

One can also read off from the distributions shown in Fig.~\ref{fig5} how
the widths of the Gaussians scale with the lattice size $L$. To a very good 
approximation we find that it is proportional to $\sqrt{L^3}$. If one
includes the sampling time dependence, the variance of $h_{\rm KS}$ scales 
like $\sqrt{L^3/t_{\rm s}}$.  As the mean value $h_{\rm KS}$ of the 
distribution $p_t(h)$ scales like $L^3$, this result confirms the
Gaussian nature of the fluctuations. Our result also has
important consequences for heavy-ion collisions. If fluctuations
are Gaussian with a dimensional scale given by the mean maximal ergodic 
Lyapunov exponent, which is found numerically to be of order
(0.5 fm)$^{-1}$ \cite{ba}, then for typical volumes and reaction times 
encountered in nuclear reactions the relative fluctuations must be very 
small, of order $\sqrt{(0.5\,{\rm fm})^4/(5\,{\rm fm})^4}=0.01$.  
This result is in agreement with a recent measurement of the fluctuations 
in relativistic heavy-ion collisions, which show that the primary 
event-by-event fluctuations in the mean value of the transverse momentum 
do not exceed 1 percent \cite{fluc1}.

Let us stress that while it is consistent to assume that the 
SU(2) gauge theory treated as a classical field theory on the lattice 
is a hyperbolic system, our positive evidence is limited.  It should 
be clear that it is impossible to exclude, by numerical calculations 
for a limited number of trajectories, that there are regions in the 
high-dimensional phase space of our lattice field theory which are 
not hyperbolic. (Then the SU(2) field on the lattice would not be an 
Anosov system.)  Also it is unproven, though highly probable, that 
the addition of the quarks will not change the picture.

\section{Conclusions}

We have shown by numerical simulations that for a two-dimensional 
billiard  the mean values for the ergodic and periodic Lyapunov
exponents and their fluctuations as a function of trajectory length 
(i.e. time) are closely related. We have derived a general relation
between their mean values and checked it numerically.
This demonstrates that we understand the relationship between 
ergodic and periodic Lyapunov exponents for the hyperbola billiard.
We have than analyzed in a similar way classical SU(2) gauge theory 
on a lattice. For all investigated properties we found good agreement 
with the expectations for a globally hyperbolic (Anosov) system. 
We conclude that for all quantities of interest which have a well-defined 
classical limit (like the growth rate of entropy after the initial 
energy deposition by hard interactions) the probability for large
fluctuations should be exponentially small.  For typical high-energy
heavy-ion collisions (Pb+Pb) such fluctuations are estimated to be
at most of the order of a few percent.

\section*{Acknowledgments}

We thank T. Guhr and M. Brack for very helpful discussions. 
B.M. acknowledges support by the Alexander von Humboldt-Stiftung 
(U.S. Senior Scientist Award) and by a grant from the U.S. Department 
of Energy (DE-FG02-96ER40495). A.S. acknowledges support by GSI and DFG.
We also acknowledge computational support by the NC Supercomputing 
Center and the Intel Corporation.

\bb{99}

\bi{NEQF} D. Boyanovsky, H.J. de Vega, R. Holman, S. Prem Kumar,
R.D. Pisarski, and J. Salgado, preprint hep-ph/9810209; 
C. Wetterich, {\sl Phys. Rev. E {\bf 56}}, 2687 (1997);
F. Cooper, S. Habib, Y. Kluger, E. Mottola, J.P. Paz, and P.R. Anderson,
{\sl Phys. Rev. D {\bf 50}}, 2848 (1994);
E. Calzetta and B.L. Hu, {\sl Phys. Rev. D {\bf 37}}, 2878 (1988).

\bi{VBE} H.T. Elze and U. Heinz, {\sl Phys. Rept. {\bf 183}}, 81 (1989);
P.F. Kelly, Q. Liu, C. Lucchesi, and C. Manuel, {\sl Phys. Rev.
D {\bf 50}}, 4209 (1994).

\bi{GM97} M. Gleiser and R.O. Ramos, {\sl Phys. Rev. D {\bf 50}}, 2441 (1994);
C. Greiner and B. M\"uller, {\sl Phys. Rev. D {\bf 55}}, 1026 (1997).

\bi{ASY99} D. B\"odeker, {\sl Phys. Lett. B {\bf 426}}, 351 (1999) and 
preprint hep-ph/9905239;
P. Arnold, D.T. Son, and L.G. Yaffe, {\sl Phys. Rev. D {\bf 59}}, 
105020 (1999).

\bi{Mat} S.G. Matinyan, G.K. Savvidy, and N.G. Ter-Arutunian Savvidy,
{\sl JETP Lett. {\bf 53}}, 421 (1981) [{\sl Pis'ma Zh. Eksp. Teor. Fiz. 
{\bf 34}}, 613 (1981)].

\bi{Sav} G.K. Savvidy, {\sl Nucl. Phys. B {\bf 246}}, 302 (1984).

\bi{ba} B. M\"uller and A. Trayanov, {\sl Phys. Rev. Lett. {\bf 68}}, 3387
(1992); T.S. Bir\'o, C. Gong, B. M\"uller, and A. Trayanov, {\sl Int. J.
Mod. Phys. C {\bf 5}}, 113 (1994).

\bi{gong1} C. Gong, {\sl Phys. Lett. B {\bf 298}}, 257 (1993).

\bi{gong2} C. Gong, {\sl Phys. Rev. D {\bf 49}}, 2642 (1994).

\bi{Biro} T.S. Bir\'o, C. Gong, and B. M\"uller, {\sl Phys. Rev. D {\bf 52}},
1260 (1995).

\bi{jb} M. Sieber, {\sl The Hyperbola Billiard: A Model for the Semiclassical 
Quantization of Chaotic Systems}, DESY preprint 91-030; M. Sieber and 
F. Steiner, {\sl Physica D {\bf 44}}, 248 (1990).

\bi{ParPol} W. Parry and M. Pollicott, {\sl Zeta Functions and the Periodic 
Orbit Structure of Hyperbolic Dynamics, Ast\'erisque} {\bf 187-188} (1990).

\bi{ga} P. Gaspard, {\sl Chaos, Scattering and Statistical Mechanics},
Cambridge University Press (1998).

\bi{wa} S. Waddington, {\sl Ann. Inst. Henri Poincar\'e C {\bf 13}}, 445 (1996).

\bi{fluc1} H. Appelsh\"auser et al. (NA49 collaboration), preprint
hep-ex/9904014.

\eb

\newpage

\newpage

\newpage

\newpage

\newpage

\newpage

\end{document}